\documentclass[reprint, amsmath,amssymb, aps, prd]{revtex4-2}

\usepackage[T1]{fontenc}
\usepackage[utf8]{inputenc}
\usepackage[english]{babel}
\usepackage{graphicx}
\usepackage{microtype}

\usepackage[colorlinks=true,
urlcolor=blue,
anchorcolor=blue,
citecolor=blue,
filecolor=blue,
linkcolor=blue,
menucolor=blue,
linktocpage=true]{hyperref}

\DeclareMathAlphabet{\mathbfsf}{OT1}{cmss}{bx}{n}

\DeclareMathOperator{\vol}{\mathrm{vol}}

\newcommand{\cI}{\mathcal{I}}

\newcommand{\cL}{\mathcal{L}}

\newcommand{\cN}{\mathcal{N}}

\newcommand{\beq}{\begin{equation}}
\newcommand{\eeq}{\end{equation}}

\newcommand{\hook}{\mathbin{\rule[.2ex]{.4em}{.03em}\rule[.2ex]{.03em}{.9ex}}}

\newcommand{\ii}{{\rm i}}
\newcommand{\e}{{\rm e}}

\newcommand{\rd}{{\rm d}}

\newcommand{\zbar}{\overline{z}}
\newcommand{\ph}[1]{\phantom{#1}}

\newcommand{\E}{{\rm E}}

\newcommand{\cone}{\alpha_1}
\newcommand{\ctwo}{\alpha_2}
\renewcommand{\j}{\varphi}

\begin{document}

\title{The Supersymmetry of Higher-Derivative Supergravity in \texorpdfstring{AdS$_4$}{AdS4} Holography}

\author{Pietro Benetti Genolini}
 \email{pietro.benettigenolini@damtp.cam.ac.uk}
\affiliation{
 Department of Applied Mathematics and Theoretical Physics, \\
University of Cambridge, Wilberforce Road, Cambridge, CB3 OWA, UK
}

\author{Paul Richmond}
 \email{paul.richmond@kcl.ac.uk}
\affiliation{
 Department of Mathematics, \\
King's College London, Strand, WC2R 2LS, UK
}

\date{\today}

\begin{abstract}
An action for the higher-derivative corrections to minimal gauged supergravity in four dimensions has been recently proposed. We demonstrate that the supersymmetric solutions of this model are those of the two-derivative action, and investigate some of their properties. In particular, we prove a formula for the renormalised on-shell action in terms of contributions from fixed points of a $U(1)$ action, and confirm that it is invariant under deformations which preserve the boundary almost contact structure.
\end{abstract}

\maketitle

\section{Introduction}
\label{sec:Intro}

Driven by the technique of supersymmetric localization, there are by now many examples of exact results for field theory observables in diverse dimensions preserving different amounts of supersymmetry.
One particularly well-studied case is that of three-dimensional $\cN=2$ field theories on curved backgrounds. Rigid supersymmetry requires that the background admits a transversely holomorphic foliation \cite{Closset:2012ru}, and fixes the dependence of supersymmetric observables on the background: in particular, the partition function is independent of deformations preserving the choice of transversely holomorphic foliation \cite{Closset:2013vra}.

Via the AdS/CFT correspondence, for those field theories which admit a holographic dual, a field theory observable has a quantum gravity counterpart. Thus, given an exact field theory observable computed via localization there is a precise prediction for a computation on the gravity side and vice versa. The simplest case is the field theory partition function, which is dual to the holographically renormalised on-shell action of the bulk gravity solution.

In practice, the situation under best computational control is in the limit in which the field theory observable is taken to be the leading contribution to the large $N$ and large 't Hooft coupling expansion, and the gravity side is classical supergravity. In order to explore the next-to-leading order contributions to field theory, it is necessary to look at higher-derivative corrections to the supergravity action.

In this note, we focus on minimal gauged supergravity in four dimensions, which describes interactions between metric and electro-magnetic field. By the AdS/CFT correspondence, its solutions are dual to the dynamics of the stress-energy tensor of a three-dimensional $\cN=2$ theory. It is generically difficult to write down the higher-derivative corrections to a supergravity theory. However, for the theory of interest here, a four-derivative action has been recently suggested \cite{Bobev:2020egg, Bobev:2021oku}. We are going to take this as our starting point, and we are going to focus on the solutions to this theory that preserve supersymmetry.

\textit{A priori}, there is no reason to believe that the two-derivative solutions would be solutions to the four-derivative equations of motion. However, remarkably this holds for this specific theory \cite{Bobev:2020egg}. In fact, we show that \textit{all} the supersymmetric solutions to the four-derivative theory are the supersymmetric solutions of the two-derivative theory. Starting from this, we find a formula for their on-shell action including the four-derivative corrections, extending the two-derivative case of \cite{BenettiGenolini:2019jdz}, and proving a conjecture in \cite{Bobev:2021oku}. One feature of this formula is that it suggests a localization theorem is at play: every supersymmetric solution admits a Killing vector $\xi$, and the on-shell action is expressed in terms of contributions from the fixed point sets of $\xi$, whether isolated (nuts) or two-dimensional (bolts).

The field-theoretic statement that rigid supersymmetry fixes the dependence of the partition function on the background is valid at finite $N$ and consequently holds at \emph{all} orders in a large $N$ expansion. In the bulk gravity dual, this translates to the requirement that the renormalised on-shell action should not depend on boundary variations that leave the transversely holomorphic foliation intact. We confirm that this holds for the four-derivative corrections and counterterms required by holographic renormalization. Indeed, demanding invariance may be a paradigm to constrain the form of six- and higher-derivative terms.

\section{Higher-derivative supergravity theory}\label{sec:Higher derivative supergravity theory}

The bosonic field content of minimal four-dimensional supergravity is a metric $g_{\mu\nu}$ and a $U(1)$ gauge field $A_\mu$. We focus on their interactions governed by the following action
\begin{align}
\label{eq:SHD}
S_{\rm HD} &= S_{2\partial} + (\cone - \ctwo) \, S_{W^2} + \ctwo \, S_{\rm GB} \, , 
\end{align}
where
\begin{align}
\label{eq:S2d}
S_{2\partial} &= - \frac{1}{16\pi G_4} \int_{Y} \big[ R + 6 L^{-2} - F_{\mu\nu} F^{\mu\nu} \big] \, \vol_g \, , \\[5pt]
\label{eq:SW2}
\begin{split}
S_{W^2} &= \int_{Y} \big[ W_{\mu\nu\rho\sigma} W^{\mu\nu\rho\sigma} - 4L^{-2} F_{\mu\nu} F^{\mu\nu} \\
& \qquad \quad + 8 F_{\mu\nu} F^{\nu\rho} F_{\rho\sigma} F^{\sigma\mu} - 2 F_{\mu\nu} F^{\mu\nu} F_{\rho\sigma} F^{\rho\sigma} \\
& \qquad \quad - 8 F_{\mu\rho} F_\nu{}^\rho R^{\mu\nu} + 2 F_{\mu\nu} F^{\mu\nu} R \\
& \qquad \quad + 8 \nabla^\mu F_{\mu\nu} \nabla^\rho F_{\rho}{}^\nu \big] \, \vol_g \, .
\end{split} \\[5pt]
\label{eq:SGB}
S_{\rm{GB}} &= \int_{Y} \big[ R_{\mu\nu\rho\sigma} R^{\mu\nu\rho\sigma} - 4 R_{\mu\nu} R^{\mu\nu} + R^2 \big] \, \vol_g \, .
\end{align}
Here $-3L^{-2}$ is the cosmological constant, $W_{\mu\nu\rho\sigma}$ is the Weyl tensor which, as with the other curvature tensors, is computed using $g$ and $F = \rd A$ is the $U(1)$ curvature.

This action is made up of three parts:\ $S_{2\partial}$, the two-derivative action, $S_{W^2}$ is a supersymmetrised version of the Weyl squared action, and $S_{\rm GB}$ is the Gauss--Bonnet action (which is topological in four dimensions). The constants $\cone$ and $\ctwo$ in front of the higher-derivative terms are arbitrary.

The action has been obtained in \cite{Bobev:2020egg, Bobev:2021oku} starting from the Weyl multiplet of four-dimensional off-shell $\cN=2$ conformal supergravity coupled to one vector multiplet and one hypermultiplet (the latter two being compensator multiplets). Then to begin to reduce to Poincaré supergravity, a number of the superconformal symmetries are gauge-fixed. This leaves us with an action involving extra superconformal fields, whose presence is required to ensure matching of the off-shell degrees of freedom. The final step is to eliminate these extra superconformal fields by solving their equations of motion in terms of $g, A$. Whilst some of these equations of motion (which can be found in \cite[(2.29)--(2.32)]{Bobev:2021oku}) are algebraic, and thus correspond to auxiliary fields, the remaining ones are generically differential equations if we include the higher-derivative corrections (that is, we take $\cone\neq \ctwo$), and they are very difficult to solve. There is an obvious solution, the ``two-derivative ansatz,'' obtained by setting $\cone=\ctwo=0$. Upon choosing these values all the equations of motion for the extra superconformal fields become algebraic and they can be easily solved. An additional peculiarity of the system of equations, though, is that the ``two-derivative ansatz'' is not only a solution of the equations with $\cone=\ctwo$, but also of the full system with arbitrary $\cone\neq \ctwo$. Thus, it is consistent to substitute it, together with the gauge-fixing conditions, in the conformal supergravity bosonic action, obtaining $S_{\rm HD}$ in \eqref{eq:SHD}.

\smallskip

Each term in \eqref{eq:SHD} is separately constructed out of superconformal invariants. Therefore, each of them is independently invariant under the supersymmetry transformations of conformal supergravity. Since the procedure of gauge-fixing and eliminating the superconformal fields using the ``two-derivative ansatz'' is consistent, we see that the final action \eqref{eq:SHD} is invariant under the supersymmetry transformations obtained from the supersymmetry transformations of conformal supergravity by gauge-fixing and substituting the ``two-derivative ansatz.'' The only relevant supersymmetry transformation for us is that of the gravitino, which is nothing but the corresponding transformations of the two-derivative minimal supergravity
\beq
\label{eq:SUSY}
\delta \psi_\mu = 2\left( \nabla_\mu \epsilon - \ii L^{-1}\, A_\mu\, \epsilon + \tfrac{L^{-1}}{2}\Gamma_\mu \, \epsilon + \tfrac{\ii}{4}F_{\nu\rho} \,\Gamma^{\nu\rho} \Gamma_\mu \epsilon \right) \, .
\eeq
The fact that the action \eqref{eq:SHD} is invariant under the supersymmetry transformations of two-derivative minimal supergravity is enforced by construction: each term is independently invariant under the supersymmetry transformations, and the ``two-derivative ansatz'' takes the first term to the two-derivative action (as the name suggests). Whilst the method just outlined leads to a supersymmetric action that includes four-derivative corrections to $S_{2\partial}$ and starts from the most general conformal supergravity action consistent with physical assumptions \cite{Bobev:2021oku}, it would be interesting to investigate whether this is the most general form while remaining in Poincaré supergravity and imposing a modified supersymmetry transformation. This is beyond the scope of this note, but we plan to return to this question in the future \cite{WIP}.

The equations of motion coming from \eqref{eq:SHD} have the form
\beq
\label{eq:EOM}
\begin{split}
0 &= E_g^{2\partial} - 16\pi G_4 (\cone - \ctwo) \, E_g^{4\partial} \, , \\
0 &= E_A^{2\partial} - 16\pi G_4  (\cone - \ctwo) \, E_A^{4\partial} \, ,
\end{split}
\eeq
where $E_g^{2\partial}$ and $E_A^{2\partial}$ are the contributions from $S_{2\partial}$, and $E_g^{4\partial}$ and $E_A^{4\partial}$ are the contributions from $S_{W^2}$ ($S_{\rm GB}$ does not contribute to the bulk equations of motion, being topological in four dimensions). Each of these terms is written in Appendix \ref{app:A}.

It is possible to show by explicit substitution that a pair $(g,A)$ for which $E_g^{2\partial}$ and $E_A^{2\partial}$ vanish also gives vanishing $E_g^{4\partial}$ and $E_A^{4\partial}$ (see Appendix \ref{app:A}) \cite{Bobev:2021oku} . Therefore, any two-derivative solution is also a four-derivative solution. Nonetheless, \textit{a priori} there could be additional solutions of the theory $S_{\rm HD}$ that are genuinely four-derivative solutions.

\section{Supersymmetric solutions}\label{sec:Supersymmetric solutions}

We are interested in the supersymmetric solutions of $S_{\rm HD}$. That is, solutions $(g,A)$ together with a non-identically zero Dirac spinor satisfying the (generalised) Killing spinor equation \eqref{eq:SUSY}: $\delta\psi_\mu = 0$. First, we recall that the latter is consistent with the two-derivative equations of motion and the Bianchi identity for the $U(1)$ gauge curvature: the integrability condition $\cI_{\rho\sigma} \sim [ \delta\psi_\rho , \delta\psi_\sigma]$ contracted with $\Gamma_\mu^{\ph{\mu}\rho\sigma}$ gives
\beq
\label{eq:Integrability}
\begin{split}
0 &= \Gamma_\mu^{\ph{\mu}\rho\sigma}\cI_{\rho\sigma} \\
&=  ({E_g^{2\partial}})_{\mu\nu}\Gamma^{\nu}\epsilon + 2 \ii (E^{2\partial}_A)_\mu \epsilon - \tfrac{\ii}{3} (\rd F)_{\nu\rho\sigma}\Gamma_\mu^{\ph{\mu}\nu\rho\sigma}\epsilon  \, .
\end{split}
\eeq

Using standard spinor bilinears techniques, it is possible to show that any supersymmetric configuration admits a Killing vector $\xi$ constructed from the spinor as $\xi = - \ii \epsilon^\dagger \Gamma_{(1)}\Gamma_5 \epsilon$. In fact, more is true, because the Killing equation implies that $\xi$ generates a symmetry of the full configuration, namely $\cL_\xi F = 0$, provided the Bianchi identity holds. At a generic point of the spacetime manifold, a non-chiral Dirac spinor generates an orthonormal frame $\{ \E^1, \E^2, \E^3, \E^4\}$. One then derives from \eqref{eq:SUSY} a number of differential equations relating the fields and the spinor bilinears expressed in terms of $\E^a$, as discussed in detail in \cite{BenettiGenolini:2019jdz} (an earlier derivation using a different technique appears in \cite{Dunajski:2010uv}). From these differential equations coming from the Killing spinor equation, without using the equations of motion, it is possible to construct the local form of the metric and gauge field. The result is
\beq
\label{eq:LocalSUSYSol}
\begin{split}
\rd s^2 &= S^2 \sin^2\theta \, \eta^2 + \frac{1}{y^4S^2\sin^2\theta}\left( \rd y^2 + 4 \e^W \rd z \rd \zbar \right) \, , \\
A &= \left( S \cos\theta + c_{\j} \right) \eta + \frac{\ii}{4} \left( \partial_z W \, \rd z - \partial_{\zbar} W \, \rd \zbar \right) \, .
\end{split}
\eeq
Here $S$ and $\theta$ are global functions on the spacetime, $\psi$ is the coordinate constructed along the orbits of $\xi$, $\eta\equiv \langle \xi, \xi \rangle_g^{-1}\xi^\flat$ is a one-form that is globally defined outside the fixed points of $\xi$, $y$ is the radial coordinate, $z, \zbar$ are local complex coordinates, $W(y,z,\zbar)$ is a local real function, and $c_{\j}$ is a real constant. There are also additional constraints among the functions, which can be derived from the supersymmetry condition
\begin{align}
\label{eq:FirstConstraint}
\frac{y}{4}\partial_y W &= 1 - \frac{1}{yS \sin^2\theta} \, , \\[5pt]
\label{eq:SecondConstraint}
\rd\eta &= 2 \left( S \sin\theta \right)^{-3} *_{\gamma} \left[ 2\cot\theta \, \rd\left(\frac{1}{y}\right) - S\, \rd\theta\right]  \, , \\[5pt]
\label{eq:ThirdConstraint}
\partial^2_{z\zbar}W &= - \e^W\left[\partial^2_{yy}W + \frac{1}{4}(\partial_yW)^2 + \frac{12\cos^2\theta}{y^4S^2\sin^4\theta} \right]
\end{align}
and $\gamma$ is
\beq
\gamma = \frac{1}{y^4}\left( \rd y^2 + 4\e^W\rd z \rd\zbar \right) \, .
\eeq
The crucial point of this analysis is that it does not require the equations of motion, only the Killing spinor equation. This is a consequence of the analysis in \cite{BenettiGenolini:2019jdz} or the reduction of the theory considered in \cite{BenettiGenolini:2018iuy}, where the necessity and sufficiency had been shown (we expand on this in Appendix \ref{app:B}). A pair $(g,A)$ solves the two-derivatives equations of motion (consistent with the integrability equation \eqref{eq:Integrability}). As already observed, every solution to the two-derivative equations of motion also solves the four-derivative equations of motion. Therefore, we conclude that all the supersymmetric solutions of the higher-derivative action \eqref{eq:SHD} are the supersymmetric solutions of the two-derivative action \eqref{eq:S2d}, and have the form \eqref{eq:LocalSUSYSol}.

\medskip

It is important at this point to make a couple of remarks on the geometry of the solution. First, notice that a solution of the Killing spinor equation has charge one under the $U(1)$ gauge field, so it generically defines a global spin$^c$ spinor. Therefore, there is no restriction on the topology of the underlying spacetime manifold, since all four-dimensional manifolds are spin$^c$. Secondly, the orbits of the Killing vector $\xi$ may close, in which case it defines a well-defined $U(1)$ isometry and we can write the four-manifold $Y$ as a circle fibration over a base $B$ with metric $\gamma$. However, it is also possible that not all orbits close, in which case we assume that the closure of the orbits of $\xi$ in the isometry group of $Y$ is a compact group, which guarantees at least a $U(1)^2$ isometry and we can approximate $\xi$ by a sequence of Killing vectors. Finally, we notice that the orthonormal frame $\E^a$ constructed from the Killing spinor could degenerate on subspaces of $Y$, where the spinor vanishes or becomes chiral. This happens precisely at the fixed points of the Killing vector $\xi$. These loci will be crucial in the next section for the evaluation of the on-shell action of the solutions.

\section{On-shell action}\label{sec:On-shell action}

The supersymmetric solutions \eqref{eq:LocalSUSYSol} are asymptotically locally anti-de Sitter, and $1/y$ has the r\^ole of a radial coordinate with $\{ y=0\}$ being the boundary of the spacetime. We then develop an expansion in $y$ assuming that the objects admit an analytic expansion in $y$ near the boundary \cite{Genolini:2016ecx}. To leading order in $y$, the result is
\beq
\begin{split}
\label{eq:AlAdSExpansion}
\rd s^2 &= \frac{\rd y^2}{y^2} + \frac{1}{y^2}\left[ \eta_{(0)}^2 +4 \e^{W_{(0)}}\rd z \rd \zbar \right] \, , \\
A &= -\theta_{(1)}\, \eta_{(0)}+ \frac{\ii}{4}\left( \partial_{z} W_{(0)}\, \rd z - \partial_{\zbar} W_{(0)}\,  \rd \zbar \right) \, ,
\end{split}
\eeq
where $W_{(0)}, \theta_{(1)}$ are functions on the boundary, and $\eta_{(0)}$ is the restriction of $\eta$ to $\partial Y$. 

\medskip

The value of the gravitational on-shell action generically diverges, but for asymptotically locally anti-de Sitter spacetime we regularize it using holographic renormalization which involves considering a cut-off spacetime $Y_\delta$ where $y\geq \delta$, adding local counterterms constructed using the induced geometry on $\partial Y_\delta$ with induced metric $h_{ij}$, and finally taking the limit $\delta \to 0$. 

For the standard action $S_{2\partial}$ in \eqref{eq:S2d}, this procedure is well-known \cite{Emparan:1999pm}, and the local counterterms include the Gibbons--Hawking--York term that imposes the Einstein equations in the bulk \cite{Gibbons:1976ue}
\beq
\label{eq:Ict2d}
\begin{split}
I_{\rm GHY} &= - \frac{1}{8\pi G_4} \int_{\partial Y_\delta} K \, \vol_h \, , \\
I_{2\partial, {\rm ct}} &= + \frac{1}{8\pi G_4} \int_{\partial Y_\delta} \left( \frac{2}{L} + \frac{L}{2} R \right) \, \vol_h \, .
\end{split}
\eeq
Here, $K_{ij}$ is the extrinsic curvature of $\partial Y_\delta$. Together with the on-shell action evaluated on the cut-off spacetime, which we denote by $I_{2\partial, {\rm bulk}}$, these give a finite quantity.

The supersymmetrised Weyl squared term $S_{W^2}$ in \eqref{eq:SW2} is more difficult to deal with. However, following \cite{Bobev:2021oku}, we observe that for a two-derivative solution, its on-shell value is greatly simplified. In fact, it can be expressed in terms of the on-shell $I_{2\partial, {\rm bulk}}$ and $I_{\rm GB, bulk}$
\beq
I_{W^2, {\rm bulk}} = - {64\pi G_4}{L^{-2}}I_{2\partial, {\rm bulk}} + I_{\rm GB, bulk} \, .
\eeq
Therefore, its evaluation is reduced to the evaluation of the other two terms.

Finally, the last term is a topological term in four dimensions that for a closed four-manifold would be proportional to the Euler characteristic, the index of the de Rham complex. Here we are instead considering its value $I_{\rm GB, bulk}$ on a manifold with a boundary $Y_\delta$, in which case the index of a complex generically receives a correction from a Chern--Simons-type form on the boundary $\partial Y_\delta$, and a correction proportional to the $\eta$ invariant of an operator on the boundary \cite{Eguchi:1980jx}. For the de Rham complex, though, the $\eta$ invariant is not relevant, and we can define the Euler characteristic by summing to $I_{\rm GB, bulk}$ the following boundary term
\beq
I_{{\rm GB}, {\rm ct}} = \int_{\partial Y_\delta} \left[ -2 K^{ij}G_{ij} + J \right] \, \vol_h \, .
\eeq
Here, as in \eqref{eq:Ict2d}, the curvature tensors have been computed using the induced metric $h_{ij}$ on $\partial Y_\delta$, $G_{ij}$ is the Einstein tensor and 
\beq
\label{eq:DefinitionJ}
\begin{split}
J_{ij} &=  \frac{1}{3} \big( 2 K K_{ik} K^{k}{}_j + K_{kl} K^{kl} K_{ij} \\
& \quad \quad \ \ - 2 K_{ik} K^{kl} K_{lj} - K^2 K_{ij} \big) \, , \\[3pt] 
J &= h^{ij} J_{ij} \, .
\end{split}
\eeq
Since $K_{ij}$ is a symmetric tensor on a three-dimensional space, it is easy to check that 
\beq
\label{eq:Simplifying}
3 \, J_{ij} = J \, h_{ij} = - 6 \det (K_{ij}) \, h_{ij} \, .
\eeq
This counterterm also guarantees that we have a well-defined boundary problem, in the sense that in the bulk it enforces the Einstein equations of motion, in the same way as the Gibbons--Hawking--York term does for the two-derivative action \cite{Bunch_1981, Myers:1987yn}. Therefore, we have
\beq
\chi(Y) \equiv 32\pi^2 \, \lim_{\delta \to 0} \left[ I_{\rm GB, bulk} + I_{{\rm GB}, {\rm ct}} \right] \, .
\eeq

Overall, we conclude that the on-shell action for all supersymmetric solutions to the higher-derivative theory can be written as
\begin{widetext}
\beq
\label{eq:IHDHoloRen}
\begin{split}
I_{\rm HD} &= \lim_{\delta \to 0} \left[ \left( 1- (\cone-\ctwo) {64\pi G_4}{L^{-2}} \right) \left( I_{2\partial, {\rm bulk}}  + I_{\rm GHY} + I_{2\partial, {\rm ct}}  \right) + \cone \left( I_{\rm GB, bulk} + I_{{\rm GB}, {\rm ct}} \right)\right] \\
&= \left( 1- (\cone-\ctwo) {64\pi G_4}{L^{-2}} \right) I_{2\partial} + 32\pi^2 \cone \, \chi(Y) \, .
\end{split}
\eeq
\end{widetext}

\medskip

The holographically renormalized on-shell action $I_{2\partial}$ of a supersymmetric solution can be written solely using geometrical data \cite{BenettiGenolini:2019jdz}. More precisely, it can be expressed in terms of contributions from the fixed loci of the Killing vector field $\xi$ constructed from the supersymmetry spinor. There are two families of those: either they are zero-dimensional nuts or two-dimensional bolts \cite{Gibbons:1979xm}. As mentioned, the fixed points of $\xi$ are those where the supersymmetry spinor $\epsilon$ becomes chiral, so they are also labelled by a sign $\pm$ representing the chirality of the spinor there. The resulting expression is
\begin{widetext}
\beq
\label{eq:I2d}
I_{2\partial}  = \frac{\pi L^2}{2G_4}\left(\sum_{\mathrm{nuts}_\mp} \pm \frac{(b_1\pm b_2)^2}{4b_1b_2} + \sum_{\mathrm{bolts}\ \Sigma_\pm} \int_{\Sigma_\pm} \left(\frac{1}{2}c_1(T\Sigma_\pm) \mp \frac{1}{4}c_1(N\Sigma_\pm)\right) \right)~,
\eeq
\end{widetext}
where $b_1, b_2$ are the weights of the rotations generated by $\xi$ on the orthogonal planes in the tangent space to the isolated nut fixed point; $T\Sigma_\pm$ and $N\Sigma_\pm$ are the tangent and normal bundle to the bolt $\Sigma_\pm$ and $c_1$ is the first Chern class of these line bundles.

The holographically renormalized on-shell action $I_{\rm HD}$ of a supersymmetric solution is expressed in terms of $I_{2\partial}$ and the Euler characteristic. The crucial property of the Euler characteristic for us is that there are a number of theorems that express it in terms of contributions from fixed point sets of actions on the manifold. In particular, the Euler characteristic of a closed manifold with an isometry is given by the sum of the Euler characteristics of each fixed point set. This still holds in the case of a manifold with a boundary, provided that the Killing vector is everywhere tangent to the boundary, as is for us (or if it is everywhere normal) \cite{Gibbons:1979xm}. Therefore, we conclude that
\beq
\chi(Y) = \sum_{\substack{{\rm fixed}\\ {\rm points}}}\chi = \sum_{\rm nuts}1 + \sum_{{\rm bolts} \ \Sigma} \int_{\Sigma} c_1(T\Sigma) \, ,
\eeq
using the Gauss--Bonnet theorem for surfaces. Inserting this in \eqref{eq:IHDHoloRen}, we find the result
\begin{widetext}
\begin{align}
I_{\rm HD} =& \frac{\pi L^2}{2G_4} \sum_{\mathrm{nuts}_\mp} \pm \frac{(b_1\pm b_2)^2}{4b_1b_2} + 32\pi^2 \,  \sum_{\mathrm{nuts}_\mp} \left[ \pm \ctwo \frac{(b_1\pm b_2)^2}{4b_1b_2} \mp \cone \frac{(b_1\mp b_2)^2}{4b_1b_2} \right]  \\
& + \frac{\pi L^2}{2G_4} \sum_{\mathrm{bolts}\ \Sigma_\pm} \int_{\Sigma_\pm} \left(\frac{1}{2}c_1(T\Sigma_\pm) \mp \frac{1}{4}c_1(N\Sigma_\pm)\right)  +32\pi^2 \sum_{\mathrm{bolts}\ \Sigma_\pm} \int_{\Sigma_\pm} \left(\frac{\cone+\ctwo}{2}c_1(T\Sigma_\pm) \pm \frac{\cone-\ctwo}{4}c_1(N\Sigma_\pm)\right)  \, , \nonumber
\end{align}
\end{widetext}
thus confirming the conjectures in \cite{Bobev:2021oku} based on a clever study of the examples. Notice that the renormalized on-shell action for a supersymmetric solution only depends on data of the isometry action of $\xi$, suggesting some sort of equivariant localization theorem.

This formula can not only be applied to the known examples to compute the corrections to the on-shell action due to higher derivatives, but it can also predict the value of the observable for other topologies, assuming that the solution exists. A number of examples can be found in \cite{BenettiGenolini:2019jdz, Bobev:2021oku}. 

\section{Supersymmetry at the boundary}\label{sec:Supersymmetry at the boundary}

Asymptotically locally anti-de Sitter supersymmetric solutions $(Y, g, A)$ induce on their conformal boundary $(M_3, \mathtt{g}, A^{(R)})$ a supersymmetric structure \cite{Klare:2012gn}. As predicted by the AdS/CFT correspondence, this structure is the same as the rigid supersymmetry constructed by coupling to non-dynamical new minimal supergravity in three dimensions \cite{Closset:2012ru}. Specifically, we identify from \eqref{eq:AlAdSExpansion} the metric on the boundary and the $U(1)_R$ background gauge field
\beq
\begin{split}
\rd s^2_3 &= \eta_{(0)}^2 +4 \e^{W_{(0)}}\rd z \rd \zbar \, , \\
A^{(R)} &= -\theta_{(1)} \, \eta_{(0)} + \frac{\ii}{4}\left( \partial_{z} W_{(0)}\, \rd z - \partial_{\zbar} W_{(0)}\,  \rd \zbar \right) \, .
\end{split}
\eeq
Geometrically, three-dimensional rigid supersymmetric backgrounds admitting two supercharges with opposite $R$ charge are manifolds with a transversely holomorphic foliation with a compatible metric, and the vector generating the foliation is Killing. Concretely, the restriction of the Killing vector field $\xi = \partial_\psi$ to the boundary is the Reeb vector field associated to the foliation, $z,\zbar$ are the coordinates on the complex leaf. In the formulation in terms of almost contact structure, the global almost contact one-form is $\eta_{(0)}$ and the expansion of \eqref{eq:SecondConstraint} to the boundary leads to the constraint
\beq
\rd\eta_{(0)} = 4\ii \e^{W_{(0)}}\theta_{(1)} \, \rd z \wedge \rd \zbar \, .
\eeq

\medskip

We shall now assume that  it is possible to consistently truncate eleven-dimensional supergravity (with its higher-derivative corrections) on a seven-manifold $X_7$ in order to obtain $S_{\rm HD}$ in \eqref{eq:SHD}. For the two-derivative action \eqref{eq:S2d}, this assumption has been proved by generalizing the Freund--Rubin background and $X_7$ being a Sasaki--Einstein manifold (see \cite{Gauntlett:2007ma} for the local uplift, and \cite{Martelli:2012sz, Toldo:2017qsh} for a careful analysis of global issues).  This procedure would also fix the coefficients $\cone, \ctwo$.

Once we make this assumption, the AdS/CFT dictionary tells us that $S_{\rm HD}$ captures universal features of three-dimensional $\cN=2$ SCFTs admitting a gravity dual, namely the dynamics of their stress-energy tensor supermultiplet. The main statement is that the partition function of the SCFT on the supersymmetric background $(M_3, \mathtt{g}, A^{(R)})$ is equal (in the large $N$ limit) to minus the logarithm of the on-shell action of the gravity dual bulk. It is known that the partition function of any three-dimensional $\cN=2$ SCFT formulated on a rigid supersymmetric background as above depends on the geometry of the background only via the choice of transversely holomorphic foliation \cite{Closset:2013vra}. That is, it is invariant under deformations $W_{(0)} \to W_{(0)} + \delta W_{(0)}, \theta_{(1)} \to \theta_{(1)} + \delta \theta_{(1)}$, where $\delta W_{(0)}(z,\zbar)$ and $\delta \theta_{(1)}(z,\zbar)$ are arbitrary global smooth functions on $M_3$ invariant under $\partial_\psi$. Thanks to AdS/CFT, this leads to an equivalent statement for the holographically renormalized on-shell action, which should be invariant under the same variations of the boundary structure.

This was proved in \cite{Genolini:2016ecx} for $I_{2\partial}$, and here we shall consider the higher-derivative corrections. As we saw in \eqref{eq:IHDHoloRen}, the higher-derivative corrections considered here (namely \eqref{eq:SHD}) are such that for supersymmetric solutions the on-shell action is simply a combination of the two-derivative on-shell action and the Euler characteristic of the bulk. Therefore, given the results of \cite{Genolini:2016ecx}, the conclusion seems to follow immediately. However, to err on the safe side, we shall now consider this explicitly.

A variation of the boundary data corresponds to a variation of the on-shell action that is necessarily a boundary term, provided the absence of boundaries or singularities in the interior. The variation $\delta I_{2\partial}$ resulting from the relevant variation of $\mathtt{g}_{ij}$ and $A^{(R)}_i$ vanishes, being exact on the base of the three-dimensional fibration. So we should simply consider the variation of $\chi$. The variation of $I_{\rm GB, bulk}+I_{\rm GB, ct}$ gives a vanishing bulk term proportional to the Lovelock tensor, and a boundary contribution that in a generic dimension has the form \cite{Davis:2002gn}
\beq
\begin{split}
T^{\rm GB}_{ij} &= - \frac{2}{\sqrt{\mathtt{g}}} \frac{\delta I_{\rm GB}}{\delta \mathtt{g}^{ij}} \\
&= \lim_{\delta\to 0} \frac{4}{\delta}  \big( 3 J_{ij} - J \, h_{ij} + 2 P_{iklj} K^{kl} \big) \, .
\end{split}
\eeq
Here, $J_{ij}$ is defined as in \eqref{eq:DefinitionJ}, whereas $P_{ijkl}$, the divergence-free part of the Riemann tensor, is 
\beq
\begin{split}
P_{ijkl} = R_{ijkl} + 2 R_{j[k} h_{l]i} - 2 R_{i[k} h_{l]j} + R h_{i[k} h_{l]j}
\end{split}
\eeq
and everything is computed using the induced metric $h_{ij}$ on $\partial Y_\delta$. However, in three dimensions $T^{\rm GB}_{ij}\equiv 0$: the first terms vanish because of \eqref{eq:Simplifying}, and $P_{ijkl}$ coincides with the Weyl tensor, which vanishes in three dimensions. This confirms that the on-shell action with higher-derivative corrections is invariant under the variations of the boundary that we are concerned with. More interestingly, notice that we may turn the argument on its head and argue that the requirement of invariance under specific variations of the boundary data imposes constraints on the form of the higher-derivative corrections.

\begin{acknowledgments}
We would like to thank Nikolay Bobev, Anthony Charles, Kiril Hristov and Valentin Reys for comments on the draft.
The work of PBG has been partially supported by the Simons Foundation, and by the STFC consolidated grants ST/P000681/1 and ST/T000694/1. PR is funded through the STFC grant ST/L000326/1.
\end{acknowledgments}

\appendix

\begin{widetext}

\section{Equations of motion}
\label{app:A}

The equations of motion coming from \eqref{eq:SHD} can be written as \eqref{eq:EOM}, where each term is \begin{align}
\left( E_g^{2\partial} \right)_{\mu\nu} =& \ R_{\mu\nu} - \tfrac{1}{2}R g_{\mu\nu} - 3 L^{-2}g_{\mu\nu} - 2 \left( F_{\mu\rho}F_{\nu}^{\ph{\nu}\rho} - \tfrac{1}{4}g_{\mu\nu} \, F_{\rho\sigma}F^{\rho\sigma} \right) \, , \\[5pt]
\left( E_A^{2\partial} \right)_\nu =& \ \nabla^\mu F_{\mu\nu} \, , \\[5pt]
\begin{split}
\left( E_g^{4\partial} \right)_{\mu\nu} = &- 2 B_{\mu\nu} - 8 L^{-2} ( F_{\mu\rho}F_\nu{}^\rho -\tfrac{1}{4} g_{\mu\nu} F_{\rho\sigma}F^{\rho\sigma} ) \\
&\ - 8 ( F_{\mu\tau}F_\nu{}^\tau F_{\rho\sigma}F^{\rho\sigma} - \tfrac{1}{8} g_{\mu\nu} (F_{\rho\sigma}F^{\rho\sigma})^2 ) + 32 (F_\mu{}^\tau F_\nu{}^\rho F_\tau{}^\sigma F_{\rho\sigma} - \tfrac{1}{8} g_{\mu\nu} F^{\lambda\rho} F_\lambda{}^\sigma F_\rho{}^\tau F_{\sigma\tau} ) \\
&\ + 4 F_{\mu\rho} F_\nu{}^\rho R + 2 ( R_{\mu\nu} - \tfrac{1}{2} g_{\mu\nu} R ) F_{\rho\sigma}F^{\rho\sigma} + 2 g_{\mu\nu} \nabla^2 [ F_{\rho\sigma}F^{\rho\sigma} ] - 2 \nabla_\mu \nabla_\nu [ F_{\rho\sigma}F^{\rho\sigma} ] \\
&\ + 4 g_{\mu\nu} F_\rho{}^\tau F_{\sigma\tau} R^{\rho\sigma} - 8 F_{\rho\mu} F_{\sigma\nu} R^{\rho\sigma} + 16 F_{\sigma(\mu} F^{\rho\sigma} R_{\nu)\rho} \\
&\ - 4 \nabla^2 [ F_\mu{}^\rho F_{\nu\rho} ] - 4 g_{\mu\nu} \nabla_\rho \nabla_\sigma [ F^{\rho\tau} F^\sigma{}_\tau ] + 8 \nabla_\sigma \nabla_{(\mu} [ F_{\nu)}{}^\rho F^\sigma{}_\rho ] \\
&\ - 4 g_{\mu\nu} \nabla^\rho F_{\rho\tau} \nabla^\sigma F_\sigma{}^\tau + 16 \nabla_{(\mu} F_{\nu)}{}^\sigma \nabla^\rho F_{\rho\sigma} + 8 \nabla^\rho F_{\rho\mu} \nabla^\sigma F_{\sigma\nu} \\
&\ - 8 g_{\mu\nu} \nabla^\rho [ F_{\rho\tau} \nabla^\sigma F_\sigma{}^\tau ] - 16 \nabla_{(\mu} [ F_{\nu)\sigma} \nabla^\rho F_\rho{}^\sigma ] - 16 \nabla_\sigma [ F_{(\mu}{}^\sigma \nabla^\rho F_{\nu)\rho} ] \\[5pt]
=& \ 4 ( E_g^{2\partial} )_{\mu\rho} ( E_g^{2\partial} )_{\nu}{}^\rho - g_{\mu\nu} ( E_g^{2\partial} )_{\rho\sigma} ( E_g^{2\partial} )^{\rho\sigma} - \tfrac{8}{3} ( E_g^{2\partial} )_{\mu\nu} ( E_g^{2\partial} )^\rho{}_\rho + \tfrac{2}{3} g_{\mu\nu} ( E_g^{2\partial} )^\rho{}_\rho ( E_g^{2\partial} )^\sigma{}_\sigma \\
&- 8 L^{-2} ( E_g^{2\partial} )_{\mu\nu} + 2 L^{-2} g_{\mu\nu} ( E_g^{2\partial} )^\rho{}_\rho + \tfrac{8}{3} F_\mu{}^\rho F_{\nu\rho} ( E_g^{2\partial} )^\sigma{}_\sigma - 8 F_{\mu\rho} F_{\nu\sigma} ( E_g^{2\partial} )^{\rho\sigma} \\
&- 2 F_{\rho\sigma} F^{\rho\sigma} ( E_g^{2\partial} )_{\mu\nu} + \tfrac{1}{3} g_{\mu\nu} F_{\rho\sigma} F^{\rho\sigma} ( E_g^{2\partial} )^\tau{}_\tau \\
& + 2 \nabla^2 ( E_g^{2\partial} )_{\mu\nu} - \tfrac{2}{3} g_{\mu\nu} \nabla^2 ( E_g^{2\partial} )^\rho{}_\rho - 4 \nabla^\rho \nabla_{(\mu} ( E_g^{2\partial} )_{\nu)\rho} + \tfrac{2}{3} \nabla_\mu \nabla_\nu ( E_g^{2\partial} )^\rho{}_\rho + 2 g_{\mu\nu} \nabla_\rho \nabla_\sigma ( E_g^{2\partial} )^{\rho\sigma} \\
&+ 8 g_{\mu\nu} \nabla^\sigma \big[ F_{\sigma\tau} ( E_A^{2\partial} )^\tau \big] - 16 \nabla_{(\mu} \big[ F_{\nu)\sigma} ( E_A^{2\partial} )^\sigma \big] + 16 \nabla_\sigma \big[ F_{(\mu}{}^\sigma ( E_A^{2\partial} )_{\nu)} \big] \\
&- 4 g_{\mu\nu} ( E_A^{2\partial} )^\tau ( E_A^{2\partial} )_\tau + 16 \nabla_{(\mu} F_{\nu)\tau} ( E_A^{2\partial} )^\tau + 8 ( E_A^{2\partial} )_\mu ( E_A^{2\partial} )_\nu 
\end{split} \\[5pt]
\begin{split}
\left( E_A^{4\partial} \right)_\nu =& \ \nabla^\mu \big[ 4 L^{-2} F_{\mu\nu} + 16 F_{\mu\rho}F^{\rho\sigma}F_{\sigma\nu} + 4 F^2 F_{\mu\nu} + 8 R_{\rho[\mu}F^\rho_{\ph{\rho}\nu]} - 2 R \, F_{\mu\nu} + 8 \nabla_{[\mu}\nabla^\rho F_{|\rho|\nu]}  \big] \\
=& \ 16 L^{-2} (E^{2\partial}_A)_\nu + 8 \nabla^\mu \big[ 4 (E^{2\partial}_g)_{\rho[\mu}F^{\rho}_{\ph{\rho}\nu]} - F_{\mu\nu}(E^{2\partial}_g)^\rho_{\ph{\rho}\rho} + 4 \nabla_{[\mu}(E^{2\partial}_A)_{\nu]} \big] \, .
\end{split}
\end{align}
Here, we have introduced the Bach tensor coming from the variation of the Weyl-squared term
\beq
\begin{split}
B_{\mu\nu} &= - 2 R_{\mu\rho}R_\nu{}^\rho + \frac{2}{3}R \, R_{\mu\nu} + \frac{1}{2} g_{\mu\nu} R_{\rho\sigma}R^{\rho\sigma} - \frac{1}{6} g_{\mu\nu} R^2 - \frac{2}{3}\nabla_\mu\nabla_\nu R - \nabla^2 R_{\mu\nu} + \frac{1}{6} g_{\mu\nu} \, \nabla^2 R + 2 \nabla_\rho\nabla_{(\mu}R_{\nu)}^{\ph{\nu)}\rho} \, .
\end{split}
\eeq
It is clear from the rewriting that if $(E^{2\partial}_g)$ and $(E^{2\partial}_A)$ vanish, then so do $(E^{4\partial}_g)$ and $(E^{4\partial}_A)$.

\section{Bilinears and equations of motion}
\label{app:B}

A supergravity solution is supersymmetric if there exists a Dirac spinor $\epsilon$ for which the gravitino variation \eqref{eq:SUSY} vanishes (here we set $L=1$):
\beq
\nabla_\mu \epsilon - \ii \, A_\mu\, \epsilon + \tfrac{1}{2}\Gamma_\mu \, \epsilon + \tfrac{\ii}{4}F_{\nu\rho} \,\Gamma^{\nu\rho} \Gamma_\mu \epsilon = 0 \, .
\eeq
At a generic point on $Y$, $\epsilon$ defines an identity structure, and we can choose to align the Killing vector $\xi$ to one of the basis vector. As pointed out in \citep{BenettiGenolini:2019jdz}, it is then possible to show from the bilinear equations that 
\beq
\xi \hook \rd *F = 0 \, , 
\eeq
which means that the Maxwell equation along the base of the fibration induced by $\xi$ is identically satisfied, that is, $\langle \E^i, E^{2\partial}_A \rangle = 0$ where $\E^i$ is any of the basis vectors orthogonal to $\xi$. It is also possible to check from the bilinear equations that $(E^{2\partial}_g)_{\mu\nu}\xi^\nu = 0$. This may also be seen from the integrability condition for the Killing spinor equation \eqref{eq:Integrability} as follows. Multiply by $\epsilon^\dagger \Gamma_5$ to obtain (assuming the Bianchi identity)
\beq
0 = - (E^{2\partial}_g)_{\mu\nu}\xi^\nu + 2 (E^{2\partial}_A)_\mu \, \epsilon^\dagger \Gamma_5 \epsilon \, .
\eeq
If we project this equation on the directions orthogonal to $\xi$, say along $K^\mu = \epsilon^\dagger \Gamma^\mu \epsilon$, then the Maxwell part vanishes, and we are left with $(E^{2\partial}_g)_{\mu\nu}K^\mu \xi^\nu = 0$.

Consider now an analogous case: multiply the integrability by $\epsilon^\dagger$
\beq
0 = (E^{2\partial}_g)_{\mu\nu}K^\nu + 2\ii (E^{2\partial}_A)_\mu \, \epsilon^\dagger \epsilon
\eeq
and now project along $\xi$. As long as $\epsilon^\dagger\epsilon \neq 0$, then we conclude that the full Maxwell equation is implied by the supersymmetry.

From this, we can use a standard analysis (see e.g \cite{Gauntlett:2002fz}): the integrability equation is now
\beq
0 = (E^{2\partial}_g)_{\mu\nu}\Gamma^\nu \epsilon \, .
\eeq
Multiply this by $(E^{2\partial}_g)_{\mu\rho}\Gamma^\rho$ to obtain
\beq
0 = (E^{2\partial}_g)_{\mu\nu}(E^{2\partial}_g)_\mu^{\ph{\mu}\nu} \quad \Rightarrow \quad 0 = (E^{2\partial}_g)_{ij}(E^{2\partial}_g)_i^{\ph{i}j} \, .
\eeq
Because of the Euclidean signature, then each $(E^{2\partial_g})_{ij}=0$ (note that there is no sum on $i$).

\end{widetext}

\bibliography{Bib_HD}

\end{document}